\documentclass[%
 aip,
 amsmath,amssymb,
 reprint,%
]{revtex4-1}

\usepackage{graphicx}
\usepackage{dcolumn}
\usepackage{bm}

\usepackage[utf8]{inputenc}
\usepackage[T1]{fontenc}
\usepackage{mathptmx}
\usepackage{etoolbox}
\usepackage{xcolor}
\usepackage{comment}
\usepackage[mathlines]{lineno}

\makeatletter
\def\@email#1#2{%
 \endgroup
 \patchcmd{\titleblock@produce}
  {\frontmatter@RRAPformat}
  {\frontmatter@RRAPformat{\produce@RRAP{*#1\href{mailto:#2}{#2}}}\frontmatter@RRAPformat}
  {}{}
}%
\makeatother
\begin{document}

\preprint{AIP/123-QED}

\title[Magnetic Stress-Strain Interaction]{A Magnetic Analog of Pressure-Strain Interaction}
\author{M.~Hasan Barbhuiya}
\author{P.~A.~Cassak}%
 \email{mhb0004@mix.wvu.edu and Paul.Cassak@mail.wvu.edu.}
\affiliation{ 
  Department of Physics and Astronomy and Center for KINETIC Plasma Physics, West Virginia University
}%


\date{\today}

\begin{abstract}
We study the evolution equation for magnetic energy density for a non-relativistic magnetized plasma in the (Lagrangian) reference frame comoving with the electron bulk velocity. Analyzing the terms that arise due to the ideal electric field, namely perpendicular electron compression and magnetic field line bending, we recast them to reveal a quantity with a functional form analogous to the often-studied pressure-strain interaction term that describes one piece of internal energy density evolution of the species in a plasma, except with the species pressure tensor replaced by the magnetic stress tensor. We dub it the ``magnetic stress-strain interaction.'' We discuss decompositions of the magnetic stress-strain interaction analogous to those used for pressure-strain interaction. These analogies facilitate the interpretation of the evolution of the various forms of energy in magnetized plasmas, and should be useful for a wide array of applications including magnetic reconnection, turbulence, collisionless shocks, and wave-particle interactions.  We display and analyze all the terms that can change magnetic energy density in the Lagrangian reference frame using a particle-in-cell simulation of magnetic reconnection.  
\end{abstract}

\maketitle


Many modern studies of fundamental processes in magnetized plasmas rely on understanding the evolution of energy and its conversion between different forms \cite{Howes17,Matthaeus20}.  While this is relatively well-understood for strongly collisional plasmas, it is challenging for collisionless or weakly collisional plasmas such as many found in space, astrophysical, and fusion plasmas.  In this brief communication, we present new insights about the time evolution of the local magnetic energy density in a magnetized plasma, specifically its evolution in the Lagrangian, {\it i.e.,} comoving, reference frame that locally moves with the bulk velocity of the electrons.  

To establish the context for this study, we briefly review results about the evolution of 
energy density in a fully-ionized net charge neutral collisionless plasma. 
The evolution equations of the bulk kinetic energy density ${\cal E}_{k,\sigma} = (1/2) m_\sigma n_\sigma u_\sigma^2$, 
internal energy density ${\cal E}_{{\rm int},\sigma} = (3/2) n_\sigma k_B T_\sigma$, and electromagnetic energy density ${\cal E}_{EM} = (E^2 + B^2)/8\pi$ are 
\begin{eqnarray}
  \frac{\partial {\cal E}_{k,\sigma}}{\partial t} + \boldsymbol{\nabla} \cdot \left( {\cal E}_{k,\sigma} {\bf u}_\sigma \right) & = & -{\bf u}_\sigma \cdot (\boldsymbol{\nabla} \cdot {\bf P}_{\sigma}) + {\bf J}_\sigma \cdot {\bf E}, \label{eq:kinengevolve} \\
\frac{\partial {\cal E}_{{\rm int},\sigma}}{\partial t} + \boldsymbol{\nabla} \cdot ({\cal E}_{{\rm int},\sigma} {\bf u}_\sigma + {\bf q}_\sigma) & = & - ({\bf P}_\sigma \cdot \boldsymbol{\nabla}) \cdot {\bf u}_\sigma, \label{eq:intengevolve} \\
\frac{\partial {\cal E}_{EM}}{\partial t} + \boldsymbol{\nabla} \cdot {\bf S} & = & - {\bf J} \cdot {\bf E} \label{eq:PoyntingTheorem}.
\end{eqnarray}
where $\sigma$ denotes the species (electron $e$ or ion $i$), $m_\sigma$ is the constituent mass, $n_\sigma = \int d^3v f_\sigma$ is the number density, $f_\sigma$ is the phase space density, ${\bf v}$ is the velocity space coordinate, ${\bf u}_\sigma = (1/n_\sigma) \int d^3v {\bf v} f_\sigma$ is the bulk flow velocity, $k_B$ is Boltzmann's constant, $T_\sigma = (2/3k_Bn_\sigma)\int d^3v (1/2) m_\sigma v_\sigma^{\prime 2} f_\sigma$ is the effective temperature, ${\bf v}_\sigma^\prime = {\bf v} - {\bf u}_\sigma$ is the peculiar velocity, ${\bf E}$ is the electric field, ${\bf B}$ is the magnetic field, ${\bf P}_\sigma$ is the pressure tensor with elements $P_{\sigma,jk} = \int d^3v m_\sigma v_{\sigma,j}^\prime v_{\sigma,k}^\prime f_\sigma$, ${\bf J}_\sigma = q_\sigma n_\sigma {\bf u}_\sigma$ is the species current density, $q_\sigma$ is the constituent charge, ${\bf q}_\sigma = \int d^3v (1/2) m_\sigma v_\sigma^{\prime 2} {\bf v}_\sigma^\prime f_\sigma$ is the vector heat flux density, 
${\bf S} = (c/4\pi) {\bf E} \times {\bf B}$ is the Poynting flux, $c$ is the speed of light, and ${\bf J} = \sum_\sigma {\bf J}_\sigma$ is the current density. 

Equations~(\ref{eq:kinengevolve})-(\ref{eq:PoyntingTheorem}) are in the form of conservation laws, where the local energy density of each kind changes in time due to a locally diverging/converging energy flux (the divergence terms on the left hand side) or through sources/sinks (the terms on the right hand side). If the equations are integrated over a volume of interest to describe the change of energy over a whole domain, Gauss' divergence theorem implies the flux terms simply become the flux through the outer surface of the domain.  If the domain being considered is closed or isolated, infinite in extent, or periodic (such as in a simulation), then the volume-integrated flux term vanishes and the total energy in each form only changes via the source/sink terms. However, we emphasize that this is only for ``global'' energy evolution; local energy conversion is impacted by the divergence of the flux terms.

We briefly discuss some of the important terms in Eqs.~(\ref{eq:kinengevolve})-(\ref{eq:PoyntingTheorem}). The so-called pressure-strain interaction is defined (with the minus sign) as
\begin{equation}
  -({\bf P}_\sigma \cdot \boldsymbol{\nabla}) \cdot {\bf u}_\sigma. \label{eq:pressurestrain}
\end{equation}
As seen in Eq.~(\ref{eq:intengevolve}), it describes the rate of change of internal energy density resulting from a bulk flow velocity strain via compressible and/or incompressible effects \cite{del_sarto_pressure_2016,yang_PRE_2017,Yang17,del_sarto_pressure_2018}.  Significant effort has been expended studying this term in space satellite \cite{Chasapis18,Zhong19,Bandyopadhyay20,bandyopadhyay_energy_2021,zhou_measurements_2021,Wang21,Burch_PoP_2023} and theoretical/numerical \cite{Hazeltine13,Sitnov18,Parashar18,Du18,yang_scale_2019,Pezzi19,song_forcebalance_2020,Du20,Pezzi21,Arro21,Fadanelli21,Yang_2022_ApJ,Cassak_PiD1_2022,Barbhuiya22,Hellinger22,Conley24} studies.  The ${\bf J}_\sigma \cdot {\bf E}$ term in Eq.~(\ref{eq:kinengevolve}) describes the local rate of conversion of energy density due to the electric field accelerating or decelerating charged particles of species $\sigma$. Adding this term up over all the species gives ${\bf J} \cdot {\bf E},$ which shows up with a minus sign in Eq.~(\ref{eq:PoyntingTheorem}); this enforces conservation of energy, {\it i.e.,} a gain in local kinetic energy density due to ${\bf J}_\sigma \cdot {\bf E}$ must be offset by an equal loss in local electromagnetic energy density and vice versa. The divergence of the Poynting flux in Eq.~(\ref{eq:PoyntingTheorem}) describes the local divergence of electromagnetic energy flux. It can be significant for magnetized collisionless plasma systems as shown by observational \cite{Welsch_2015_PNASJ,Nishimura_2010_JGR,Volwerk_1996_JGR,Payne_2020_JGR,Genestreti_2018_GRL} and theoretical/numerical \cite{Liu_2022_CommPhys, Payne_2021_PoP, Payne_JGR_2024,Treumann&Baumjohann_2017_AnnGeoPhys, Treumann&Baumjohann_2017_arxiv} studies.


Equations~(\ref{eq:kinengevolve})-(\ref{eq:PoyntingTheorem}) are written in the (Eulerian) stationary ``laboratory'' reference frame which is at rest and the plasma and electromagnetic field energy move through it. However, at a point of interest in the Eulerian reference frame, the convection of a plasma or field from another point due to the flow of the plasma can change the energy density.  Convection does not represent a genuine change in the properties of the plasma, so the study of energy evolution is preferably carried out in the reference frame comoving with the plasma, the Lagrangian reference frame.  (See, {\it e.g.}, Ref.~\cite{Cassak_FirstLaw_2023,Tenbarge24}.)

It is not uncommon to write equations~(\ref{eq:kinengevolve}) and (\ref{eq:intengevolve}) in the Lagrangian frame. It is less common to see Poynting's theorem [Eq.~(\ref{eq:PoyntingTheorem})] written in the Lagrangian frame (see Ref.~\cite{Treumann&Baumjohann_2017_arxiv} for one example). Notably, Zenitani and coauthors \cite{Zenitani11} discussed 
the rate of energy density conversion due to acceleration of charged particles in the reference frame comoving with the bulk velocity of the electrons.  The so-called ``Zenitani parameter'' was introduced as a means to identify where non-ideal electron-scale physics takes place, such as the electron diffusion region of collisionless magnetic reconnection.  
It was originally thought to describe
dissipation \cite{Zenitani11}, but  
later it was recognized that it may include both reversible and irreversible contributions, particularly in weakly collisional plasmas 
\cite{Cassak17}.

In this study, we 
analyze 
the local magnetic energy evolution in the Lagrangian 
reference frame comoving with the bulk velocity of electrons.  By breaking the electric field into ideal and non-ideal contributions, we show the contribution due to the ideal electric field is described by terms analogous to those in ideal-magnetohydrodynamics (ideal-MHD) describing perpendicular compression and magnetic field line bending, but 
due to bulk electron motion. We recast these two terms to reveal a quantity analogous to the pressure-strain interaction term described in Eq.~(\ref{eq:pressurestrain}), but with the magnetic stress tensor ${\bf T}_M$ replacing the pressure tensor of the species, where the $jk$th element is given by $T_{M,jk} = (1/4\pi) [B_j B_k - (1/2) \delta_{jk} B^2]$ and $\delta_{jk}$ is the Kronecker delta.  We call the magnetic analog to the pressure-strain interaction the ``magnetic stress-strain interaction.''  
This term does not represent new physics concerning the evolution of magnetic energy; rather
it identifies a symmetry in the energy evolution and conversion between the different forms of energy which, given the broad interest in the pressure-strain interaction, may facilitate insight and understanding of the physical processes during energy conversion.

We first write Poynting's theorem in Eq.~(\ref{eq:PoyntingTheorem}) in the reference frame comoving with the electrons in the non-relativistic limit.  We start by writing the electric field as ${\bf E} = - {\bf u}_e \times {\bf B}/c + {\bf E}^\prime$, where $-{\bf u}_e \times {\bf B}/c$ is the convective electric field and would be the whole electric field if the plasma were ideal with the electrons frozen-in to the magnetic field, and ${\bf E}^\prime$ is the non-ideal electric field.  Then, ignoring the electric energy density because we are in the non-relativistic limit, Eq.~(\ref{eq:PoyntingTheorem}) becomes
(see also \cite{Treumann&Baumjohann_2017_arxiv}) 
\begin{equation}
    \frac{\partial {\cal E}_{M}}{\partial t} + \frac{c}{4 \pi} \boldsymbol{\nabla} \cdot \left[ \left( {\bf E}^\prime - \frac{{\bf u}_e \times {\bf B}}{c} \right) \times {\bf B} \right] = - {\bf J} \cdot \left( {\bf E}^\prime - \frac{{\bf u}_e \times {\bf B}}{c} \right) \label{eq:PoyntingTheoremComoving}.
\end{equation}
where the local magnetic energy density is ${\cal E}_M = B^2 / 8 \pi$.
Simplifying Eq.~(\ref{eq:PoyntingTheoremComoving}) by gathering terms related to ${\bf E}^\prime$, using Amp\`ere's law ${\bf J} = (c/4\pi) \boldsymbol{\nabla} \times {\bf B}$ to eliminate ${\bf J}$ in the ideal term, and using vector identities to gather terms gives the magnetic energy density evolution equation as
\begin{eqnarray}
    \frac{\partial {\cal E}_M}{\partial t} & + & ({\bf u}_{e,\perp} \cdot \boldsymbol{\nabla}) {\cal E}_M = -\frac{B^2}{4\pi} (\boldsymbol{\nabla} \cdot {\bf u}_{e,\perp}) \nonumber \\ & & + \frac{1}{4\pi} {\bf B} \cdot \left[({\bf B} \cdot \boldsymbol{\nabla}) {\bf u}_{e,\perp} \right] - \boldsymbol{\nabla} \cdot {\bf S}^\prime - {\bf J} \cdot {\bf E}^\prime, \label{eq:demdtphysics}
\end{eqnarray}
where ${\bf S}^\prime = (c/4\pi) {\bf E}^\prime \times {\bf B}$ is the non-ideal Poynting flux based solely on the non-ideal electric field ${\bf E}^\prime$, and ${\bf u}_{e,\perp} = {\bf u}_e - {\bf \hat{b}} ({\bf \hat{b}} \cdot {\bf u}_e)$ is the bulk electron velocity perpendicular to the magnetic field and ${\bf \hat{b}} = {\bf B}/B$ is the unit vector in the direction of the magnetic field. The component of ${\bf u}_e$ parallel to the magnetic field does not enter the convective electric field, and therefore cannot contribute to the magnetic energy density equation.

The form of Eq.~(\ref{eq:demdtphysics}) facilitates the identification of the physical processes at play.  The first term on the left-hand side, $\partial {\cal E}_M/\partial t$, quantifies the local change in magnetic energy density in the stationary (Eulerian) reference frame.  The second term on the left-hand side, $({\bf u}_{e,\perp} \cdot \boldsymbol{\nabla}) {\cal E}_M$, describes the rate of change of magnetic energy density due to convection of magnetic energy density by electrons perpendicular to the magnetic field.  These two terms together describe the rate of magnetic energy density conversion in the Lagrangian reference frame comoving with the perpendicular part of the bulk electron velocity ${\bf u}_{e,\perp}$.

Consequently, the four terms on the right-hand side describe mechanisms by which the local magnetic energy density changes in the reference frame comoving with the perpendicular part of the bulk electron velocity.  The first term on the right-hand side, $-(B^2/4\pi) (\boldsymbol{\nabla} \cdot {\bf u}_{e,\perp})$, is the rate of change of magnetic energy density due to compression of electron bulk flow perpendicular to the magnetic field.  The second term on the right-hand side, $(1/4\pi) {\bf B} \cdot \left[({\bf B} \cdot \boldsymbol{\nabla}) {\bf u}_{e,\perp} \right]$, is the rate of change of magnetic energy density due to magnetic field bending by the electrons.  Physically, the bending term is non-zero if the electron bulk flow perpendicular to the magnetic field changes along the direction of the magnetic field. Both of these terms are due to the ideal electric field and are analogous to the terms that arise in ideal-MHD, except the perpendicular electron bulk velocity ${\bf u}_{e,\perp}$ plays the role of the single-fluid bulk velocity in ideal-MHD. We note this analogy persists here despite the fact that we are not making any assumptions about the validity of ideal-MHD or Hall-MHD in our analysis.

The third term on the right-hand side, $-\boldsymbol{\nabla} \cdot {\bf S}^\prime$, is the rate of change of magnetic energy density due to the divergence of the non-ideal Poynting flux \cite{Treumann&Baumjohann_2017_arxiv}.  This term describes the transport of magnetic energy density rather than a pure source/sink of it, as integrating it over the entire volume of a system in question results solely in the Poynting flux across the boundary of the system.

Finally, the fourth term on the right-hand side, $-{\bf J} \cdot {\bf E}^\prime$, is the local rate of change of magnetic energy density due to the acceleration of charged particles by the non-ideal electric field \cite{Treumann&Baumjohann_2017_arxiv}.  
The 
Zenitani parameter \cite{Zenitani11} is defined 
as $D_e = \gamma_e [{\bf J} \cdot ({\bf E} + {\bf u}_e \times {\bf B}/c) - \rho_c({\bf u}_e \cdot {\bf E})],$ where $\gamma_e = (1 - u_e^2/c^2)^{-1/2}$ is the relativistic gamma factor for electrons and $\rho_c = \sum_\sigma q_\sigma n_\sigma$ is the net charge density. 
In the non-relativistic quasineutral limit, $\gamma_e \simeq 1$ and $\rho_c \simeq 0$, so the Zenitani parameter 
is approximately given by 
${\bf J} \cdot {\bf E}^\prime$.
The appearance of the Zenitani parameter as a source term in Eq.~(\ref{eq:demdtphysics}) is consistent with the expectations in Ref.~\cite{Zenitani11}.

We revisit the power density due to the ideal electric field. 
By direct calculation, we find the first two terms on the right hand side of Eq.~(\ref{eq:demdtphysics}) are 
equivalent to
\begin{equation}
     - {\cal E}_M (\boldsymbol{\nabla} \cdot {\bf u}_{e,\perp}) + ({\bf T}_{M} \cdot \boldsymbol{\nabla}) \cdot {\bf u}_{e,\perp}, \label{eq:magstressstrain}
\end{equation}
because using the definition of ${\bf T}_M$ reveals
\begin{equation}
     ({\bf T}_{M} \cdot \boldsymbol{\nabla}) \cdot {\bf u}_{e,\perp} = - {\cal E}_M (\boldsymbol{\nabla} \cdot {\bf u}_{e,\perp}) + \frac{1}{4\pi} {\bf B} \cdot[({\bf B} \cdot \boldsymbol{\nabla}) {\bf u}_{e,\perp}]. \label{eq:tmdef}
\end{equation}
The left-hand side of Eq.~(\ref{eq:tmdef}) has the same form as the pressure-strain interaction in Eq.~(\ref{eq:pressurestrain}), except that the magnetic stress tensor plays the role of the pressure tensor and the strain rate tensor is solely for the bulk flow perpendicular to the magnetic field instead of the total velocity ${\bf u}_e$.  We therefore dub $({\bf T}_{M} \cdot \boldsymbol{\nabla}) \cdot {\bf u}_{e,\perp}$ the ``magnetic stress-strain interaction'' (without a minus sign). From Eq.~(\ref{eq:tmdef}),
the physical interpretation of the magnetic stress-strain interaction is the combined effects of perpendicular electron compression and magnetic field line bending.

The pressure-strain interaction $-({\bf P}_\sigma \cdot \boldsymbol{\nabla}) \cdot {\bf u}_{\sigma}$ is commonly decomposed in ways that elucidate the physical effects underlying it, so we discuss analogous
decompositions of $({\bf T}_M \cdot \boldsymbol{\nabla}) \cdot {\bf u}_{e\perp}$. 
First,  the compressional part of pressure-strain interaction is given by \cite{Yang17} ${\cal P}_\sigma (\boldsymbol{\nabla} \cdot {\bf u}_\sigma)$, where ${\cal P}_\sigma = (1/3) {\rm tr} ({\bf P}_\sigma)$ is the effective (scalar) pressure and tr denotes the trace.  Its magnetic counterpart is ${\cal T}_M (\boldsymbol{\nabla} \cdot {\bf u}_{e,\perp})$, where we define the effective (scalar) magnetic stress as ${\cal T}_M = (1/3) T_{M,jj} = (1/3) (1/4\pi) [B^2 -(1/2) \delta_{jj}B^2] 
= -(1/3) {\cal E}_M$.  Then, the compressible part of the magnetic stress-strain interaction is
\begin{equation}
    {\cal T}_M (\boldsymbol{\nabla} \cdot {\bf u}_{e,\perp}) = -\frac{1}{3} {\cal E}_M (\boldsymbol{\nabla} \cdot {\bf u}_{e,\perp}).
\end{equation}

Next, the term that describes the net effect of converging or diverging flow for pressure-strain interaction is called \cite{Cassak_PiD1_2022} ${\rm PDU}_\sigma$. We dub the magnetic stress counterpart as ${\rm TDU}$, defined in rectangular coordinates as
\begin{equation}
{\rm TDU} = T_{M,xx} \frac{\partial u_{e,\perp,x}}{\partial x} + T_{M,yy} \frac{\partial u_{e,\perp,y}}{\partial y} + T_{M,zz} \frac{\partial u_{e,\perp,z}}{\partial z}.
\end{equation}
Using the definition of the diagonal elements of ${\bf T}_M$, a brief derivation yields
\begin{eqnarray}
{\rm TDU} & = & \frac{1}{4\pi} \left(B_x^2 \frac{\partial u_{e,\perp,x}}{\partial x} + B_y^2 \frac{\partial u_{e,\perp,y}}{\partial y} + B_z^2 \frac{\partial u_{e,\perp,z}}{\partial z} \right) \nonumber \\ & & -  {\cal E}_{M} (\boldsymbol{\nabla} \cdot {\bf u}_{e,\perp}).
\end{eqnarray}

The term that isolates the incompressible portion due to normal flow in the pressure-strain interaction was called \cite{Cassak_PiD1_2022} ${\rm Pi-D}_{{\rm normal},\sigma} = {\rm PDU}_\sigma - P_\sigma (\boldsymbol{\nabla} \cdot {\bf u}_{\sigma})$. We define the magnetic counterpart describing the incompressible portion due to normal flow as
\begin{eqnarray}
 {\rm T}{\cal D}_{{\rm normal}} & = & {\rm TDU} - {\cal T}_M (\boldsymbol{\nabla} \cdot {\bf u}_{e,\perp})
 \\
 & = & \frac{1}{4\pi} \left(B_x^2 \frac{\partial u_{e,\perp,x}}{\partial x} + B_y^2 \frac{\partial u_{e,\perp,y}}{\partial y} + B_z^2 \frac{\partial u_{e,\perp,z}}{\partial z} \right) \nonumber \\ & & - \frac{2}{3} {\cal E}_{M} (\boldsymbol{\nabla} \cdot {\bf u}_{e,\perp}),
\end{eqnarray}
since ${\cal T}_M = - (1/3) {\cal E}_M$.  The term dubbed ${\rm Pi-D}_{{\rm shear}}$ in Ref.~\cite{Cassak_PiD1_2022} internal energy density conversion due to purely sheared flow.  We refer to the magnetic analog of this as $T {\cal D}_{{\rm shear}}$, which in rectangular coordinates is
\begin{equation}
T {\cal D}_{{\rm shear}} = \frac{1}{4 \pi} \sum_{j, k \neq j} B_j B_k \frac{\partial u_{e,\perp,j}}{\partial r_k}.
\end{equation}
Finally, the analog of the incompressible part of pressure-strain interaction \cite{Yang17} ${\rm Pi-D}_\sigma = {\rm Pi-D}_{{\rm normal},\sigma} + {\rm Pi-D}_{{\rm shear},\sigma}$ for the magnetic stress-strain interaction is
\begin{equation}
T {\cal D} = \frac{1}{4 \pi} {\bf B} \cdot \left[ ({\bf B} \cdot \boldsymbol{\nabla})  {\bf u}_{e,\perp} \right] - \frac{2}{3} {\cal E}_{M} (\boldsymbol{\nabla} \cdot {\bf u}_{e,\perp}).
\end{equation}




We now present the terms in the magnetic energy density evolution equation [Eq.~(\ref{eq:demdtphysics})] using particle-in-cell simulations of magnetic reconnection. The simulations are identical to those in Ref.~\cite{Barbhuiya22} and are thoroughly described therein. The simulation data are 
presented only for 
the lower current sheet at time $t = 13 \ \Omega_{ci0}^{-1}$ with black solid lines denoting in-plane projections of magnetic field lines, 
where $\Omega_{ci0}$ is the ion cyclotron frequency based on the initial asymptotic magnetic field $B_0$. 
This time is when the reconnection rate is rising most rapidly during the nonlinear growth phase, 
so the system is not in steady-state. 
The plots are centered at the X-line $(x_0,y_0)$, and
all distances are given in units of the ion inertial scale $d_{i0}$ based on the difference between the maximum initial number density and the asymptotic value.
To minimize particle noise in the simulation,
we smooth the raw simulation data over a width of five cells, followed by calculating the desired
spatial or temporal derivatives, and finally, smooth the results 
over five cells. This smoothing approach is chosen after testing various options for the number of cells to smooth over and the number of 
recursions, ensuring that the signal structure remains largely unaffected by the smoothing. All power densities are given in units of $(B_0^2/4\pi) \Omega_{ci0}$.

\begin{figure}
    \centering
    \includegraphics[width=1\linewidth]{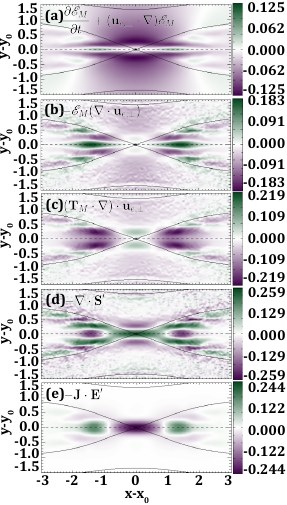}
    \caption{Particle-in-cell simulation results of the terms in the magnetic energy density evolution equation [Eq.~(\ref{eq:demdtphysics})] during magnetic reconnection.
    (a) 
    The time rate of change of magnetic energy density $\mathcal{E}_M$ in the Lagrangian frame comoving with $\mathbf{u}_{e,\perp}$, 
    (b) electron perpendicular compression $- {\cal E}_M (\boldsymbol{\nabla} \cdot {\bf u}_{e,\perp})$, (c) magnetic stress-strain interaction $({\bf T}_{M} \cdot \boldsymbol{\nabla}) \cdot {\bf u}_{e,\perp}$, (d) negative of the divergence of the non-ideal Poynting flux $-\boldsymbol{\nabla} \cdot {\bf S}^\prime$, and (e) the power density by the non-ideal electric field $-{\bf J} \cdot {\bf E}^\prime$.}
    \label{fig:E_mag_evol_terms}
\end{figure}

Panel (a) of Fig.~\ref{fig:E_mag_evol_terms} displays the left-hand side of Eq.~(\ref{eq:demdtphysics}) that denotes the time rate of change of
$\mathcal{E}_M$ in the Lagrangian frame comoving with $\mathbf{u}_{e,\perp}$.  We see that the magnetic energy density decreases (purple) with time in the Lagrangian frame upstream of the X-line and there is a weak increase (green) along the horizontal neutral line (the dashed line) downstream of the X-line. Since the reconnection process converts magnetic energy to energy in the plasma, a decrease of magnetic energy in the Lagrangian frame is expected.  The positive signal along the outflow region of the neutral line is because the reconnected magnetic field $B_y$ increases with distance in the outflow direction from the X-line while the reconnecting magnetic field $B_x = 0$, so ${\cal E}_M$ increases in the reference frame moving with the electrons. 

We now examine the terms on the right-hand side of Eq.~(\ref{eq:demdtphysics}).
Figs.~\ref{fig:E_mag_evol_terms}(b)-(e) contain the perpendicular compression term $-{\cal E}_M (\boldsymbol{\nabla} \cdot {\bf u}_{e,\perp})$, 
the magnetic stress-strain interaction $({\bf T}_{M} \cdot \boldsymbol{\nabla}) \cdot {\bf u}_{e,\perp}$, 
the negative divergence of the non-ideal Poynting flux 
$-\boldsymbol{\nabla} \cdot {\bf S}^\prime$, and
the power density by the non-ideal electric field $-{\bf J} \cdot {\bf E}^\prime$, respectively.  We consider the action of these terms in a few key regions for the reconnection process.

Near the X-line, within $|x-x_0|<1$ and $|y-y_0|<0.3$, perpendicular compression [in (b)] and magnetic stress-strain interaction [in (c)] are small because they are proportional to the magnetic energy and magnetic stress, respectively, which are small because the magnetic field is weak near the X-line. As known from previous work \cite{Zenitani11}, there is a strong $-{\bf J} \cdot {\bf E}^\prime$ [in (e)] in the inner electron diffusion region where the (non-ideal) reconnection electric field ${\bf E}^\prime$ in the $-z$ direction accelerates electrons in the $z$ direction.  Since the magnetic energy near the X-line does not change much in time, the non-ideal Poynting flux [in (d)] essentially balances $-{\bf J} \cdot {\bf E}^\prime$.  Thus, ${\bf S}^\prime$ converges near the X-line to provide the energy necessary for the reconnection electric field to accelerate the electrons.  Physically, ${\bf E}^\prime$ is in the $-z$ direction while $B_x >0$ for $y-y_0 > 0$ and $B_x < 0$ for $y-y_0 < 0$, so ${\bf E}^\prime \times {\bf B}$ is towards the neutral line both above and below the X-line.

In the outflow region around $1<|x-x_0|<2$, there is a small region of magnetic energy density increase near the neutral line, and otherwise the magnetic energy density is mostly decreasing [panel (a)]. The increase in magnetic energy is caused by perpendicular compression [panel (b), and it also contributes to the magnetic stress-strain interaction in panel (c) from Eq.~(\ref{eq:tmdef})]. Near the neutral line, magnetic field lines are strongly kinked, and the $y$ component of ${\bf u}_{e,\perp}$ points towards the neutral line, thus giving strong compression $-{\cal E}_M (\boldsymbol{\nabla} \cdot {\bf u}_{e,\perp})$, which leads to the increase in ${\cal E}_M$.  We note there is also a positive $-{\bf J} \cdot {\bf E}^\prime$ in this region because of a non-zero divergence of the electron pressure tensor \cite{Hesse08,Egedal16}. Since the sign of $E_z^\prime$ flips to positive in this region, $\boldsymbol{\nabla} \cdot {\bf S}^\prime$ is positive which nearly balances $-{\bf J} \cdot {\bf E}^\prime$, so the non-ideal electric field is not the cause of the change in ${\cal E}_M$ at this region.
The decrease in ${\cal E}_M$ away from the neutral line in the exhaust is due to magnetic field line bending. The magnetic field lines are strongly kinked, and they lose energy as they straighten. This appears in the magnetic bending term in Eq.~(\ref{eq:demdtphysics}).  In the language of the decomposition of the magnetic stress-strain interaction, the increase in ${\cal E}_M$ near the neutral line is captured as a negative
${\rm T}{\cal D}_{{\rm normal}}$ term. Away from the neutral line,
the bending is captured as a negative $T {\cal D}_{{\rm shear}}$. Importantly, both of these describe changes to the magnetic energy via incompressible effects.  This is a reasonable result because
the Alfv\'enic and whistler wave-like effect driving the outflow \cite{Mandt94,rogers01a,Drake07} are incompressible.


We next treat the region around the upstream edge of the electron diffusion region, $|x-x_0|<1$ and $0.3 < |y-y_0|<0.5$. The dominant term is $\boldsymbol{\nabla} \cdot {\bf S}^\prime$, which arises because the non-ideal reconnection electric field $E_z^\prime$ is reasonably uniform, but the strength of the reconnecting magnetic field $B_x$ decreases approaching the X-line. This leads to a converging ${\bf S}^\prime$ associated with the decrease in magnetic energy density in the Lagrangian frame.
Further upstream, outside the electron diffusion region, the magnetic energy density decreases in the Lagrangian frame [panel (a)]; this is dominated by
a decrease in $\mathcal{E}_M$ in the Eulerian frame, \textit{i.e.,} $\partial \mathcal{E}_M/\partial t<0$ (not shown). At the time slice plotted here,
the system is not in the steady-state, so the magnetic field energy density decreases locally due to the magnetic field getting pulled in toward the X-line as the reconnection process speeds up.


Finally, we make
two observations.  First, the non-ideal terms [in (d) and (e)] are predominantly localized to the electron diffusion region and within an electron inertial length size surrounding the separatrix.  This is consistent with the electron frozen-in condition breaking down at sub-electron inertial scales in anti-parallel magnetic reconnection so that ${\bf E}^\prime \neq 0$.  Second, 
although the data presented here is from a time slice when the time evolution is the most rapid, we do not expect significant differences in the structure of these plots near the electron diffusion region
in the steady-state because these plots are in the Lagrangian reference frame.  In the steady-state, $\partial/\partial t = 0$, but the convection term ${\bf u}_{e,\perp} \cdot \boldsymbol{\nabla}$ remains significant. 

The reconnection process is used as an example, but we expect the analysis performed here and its interpretation could be useful in other fundamental processes in magnetized plasmas, including turbulence, collisionless shocks, and wave-particle interactions.

\begin{acknowledgments}
This work was partially motivated by PAC's participation in the Magnetospheric Multiscale (MMS) 10th Community Meeting. We gratefully acknowledge support from NASA grants 80NSSC24K0172 and 80NSSC23K0409, NSF grant PHY-2308669, and DOE grant DE-SC0020294. This research used resources of the National Energy Research Scientific Computing Center (NERSC), a U.S. Department of Energy Office of Science User Facility located at Lawrence Berkeley National Laboratory, operated under Contract No. DE-AC02-05CH11231 using NERSC award FES-ERCAP0027083.
\end{acknowledgments}

\section*{Data Availability Statement}


The data that support the findings of this study are openly available in Zenodo at https://doi.org/10.5281/zenodo .8147803. 
  
\bibliography{magstressstrain}

\end{document}